\documentclass[a4paper, 11pt]{article}
\usepackage{graphicx} 
\graphicspath{{img/}}
\usepackage[margin=2cm]{geometry}
\usepackage{diagbox}
\usepackage{booktabs}
\usepackage{xcolor}
\usepackage[sorting=none]{biblatex}
\bibliography{./bibliography}
\usepackage{hyperref}
\usepackage{subcaption}
\usepackage{siunitx}
\usepackage{authblk}

\title{GAMBAS - Fast Beam Arrangement Selection for Proton Therapy using a Nearest Neighbour Model}

\newbox{\myorcidaffilbox}
\sbox{\myorcidaffilbox}{\large\includegraphics[height=1.7ex]{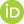}}
\newcommand{\orcidaffil}[1]{%
  \href{https://orcid.org/#1}{\usebox{\myorcidaffilbox}\,#1}}

\author[1]{Renato Bellotti \orcidaffil{0000-0002-2702-0437}}
\author[1]{Nicola Bizzocchi}
\author[1]{Antony J.\ Lomax}
\author[2]{Andreas Adelmann \orcidaffil{0000-0002-7230-7007}}
\author[1,3,4]{Damien C.\ Weber \orcidaffil{0000-0003-1166-8236}}
\author[1,*]{Jan Hrbacek}

\affil[1]{Center for Proton Therapy, Paul Scherrer Institut}
\affil[2]{Accelerator Modelling and Advanced Simulations group, Paul Scherrer Institut}
\affil[3]{Department of Radiation Oncology, University Hospital of Zurich, USZ, Zurich}
\affil[4]{Department of Radiation Oncology, University Hospital of Bern, Inselspital, Bern}

\affil[*]{Corresponding author: jan.hrbacek@psi.ch, Forschungsstrasse 111, 5232 Villigen PSI, Switzerland }

\begin{document}

\maketitle

\begin{abstract}
    \textbf{Purpose} Beam angle selection is critical in proton therapy treatment planning, yet automated approaches remain underexplored. This study presents and evaluates GAMBAS, a novel, fast machine learning model for automatic beam angle selection.
    
    \textbf{Methods} The model extracts a predefined set of anatomical features from a patient’s CT and structure contours. Using these features, it identifies the most similar patient from a training database and suggests that patient's beam arrangement. A retrospective study with 19 patients was conducted, comparing this model's suggestions to human planners' choices and randomly selected beam arrangements from the training dataset. An expert treatment planner evaluated the plans on quality (scale 1-5), ranked them, and guessed the method used.
    
    \textbf{Results} The number of acceptable (score $\geq 4$) plans was comparable between human-chosen $17~(89\%)$ and model-selected $16~(84\%)$ beam arrangements. The fully automatic treatment planning took between $4 - 7$~min (mean $5$~min).
    
    \textbf{Conclusion} The model produces beam arrangements of comparable quality to those chosen by human planners, demonstrating its potential as a fast tool for quality assurance and patient selection, although it is not yet ready for clinical use.
\end{abstract}

\section{Introduction}
Beam angle selection (BAS) is critical for achieving high-quality treatment plans for proton therapy. The beam arrangement drives where the entry-path dose is deposited, setting the boundaries for the subsequent spot weight optimisation. Despite its importance, literature on automatic BAS for intensity-modulated proton therapy (IMPT) is limited.

Cao et al.\ \cite{cao_reflections_2022} found only $5$ papers discussing automatic BAS for proton therapy, suggesting that most clinics select their beam arrangements (BAs) either manually or using template-based solutions. Most of the BAS literature targets photon therapy. However, proton treatments feature unique challenges such as range-robustness and biological effects that prohibit direct transfer of methods~\cite{cao_reflections_2022} from photon therapy.

Most of the methods reported by Cao et al.~\cite{cao_reflections_2022} are based on the same principle. The BAS algorithm suggests candidate BAs whose quality is assessed, typically based on dosimetric metrics. Then, the next set of candidate solutions is suggested. For each candidate solution, automatic spot weight optimisation and a dose calculation need to be performed, which raises the computational cost so much that it renders the automatic BAS algorithms too costly for clinical application.

Several approaches have been explored to accelerate automatic BAS. Dias et al.~\cite{dias_genetic_2014} circumvent the spot weight optimisation and dose calculation by training a neural network to predict the fitness of a BA only based on the angles. Subsequently, the neural network guides a genetic optimisation algorithm. Kaderka et al.~\cite{kaderka_toward_2022} abandon the path of optimisation altogether by framing the BAS as a multi-label classification problem that is solved by a neural network. Their network predicts the full BA based on the CT image and the structure contours for a liver case. The training requires $21$~hours on an NVIDIA 2080 graphics processing unit (GPU), but the prediction of a BA for a new patient takes only $2$~minutes. Despite being the fastest method and thus most suited for real-time applications, the neural network approach by Kaderka et al.~\cite{kaderka_toward_2022} require large datasets of similar patients, which are not available in smaller clinics.

We aim to resolve this shortcoming by introducing the Geometry-based Automatic Model for Beam Arrangement Selection (GAMBAS). GAMBAS calculates a set of geometric anatomical features that allow similarity comparisons between patients, instead of learning how to find a good beam arrangement for a given anatomy. The BAS problem is then reduced to the easier problem of identifying the most similar patient in the database of previously treated patients and suggesting the BA that was used to treat that patient.

By principle, GAMBAS BAs should be implicitly robust with respect to, including but not limited to, daily anatomic variations, range uncertainties and patient positioning uncertainties because each GAMBAS BAs have already been used to treat similar patients. Due to the model's simplicity, the training can be completed in minutes without requiring GPUs and the training set can be augmented without retraining. Further, GAMBAS is designed to support a wide range of brain and eye tumours. This capability is crucial for application at our institute as we are referred a diverse cohort of patients, most of them presenting with tumours in both regions. Finally, GAMBAS' output is a list of training patients sorted by similarity, which renders the predictions interpretable. Additional information from the treatment of the most similar patient(s) can be displayed to provide more context to clinicians.

GAMBAS is part of the open source autoplanning package developed at our institute \cite{Bellotti2023}. The automatic spot weight optimisation algorithm JulianA has already been validated, but requires a beam arrangement as input. GAMBAS fills this gap, and the combined GAMBAS-JulianA system establishes a novel fast and fully automatic treatment planning method.

This work analyses GAMBAS in a retrospective planning study and provides evidence that it produces clinically acceptable beam arrangements for a wide range of patients within minutes.

\section{Materials and Methods}
The main contribution of this work is the GAMBAS model for automatic fast and interpretable BAS. It aims to provide a clinically acceptable BA in a fast way, rather than obtaining the best-possible solution. Acceptable in this context means that the resulting plan could be used for patient treatment, i.\ e.\ it complies with our clinical protocol regarding target coverage, organ-at-risk (OAR) sparing and normal tissue sparing.

GAMBAS is a nearest neighbour model \cite{Cover1967-la}. It summarises the patient's anatomy based on a set of hard-coded features that are detailed in the following paragraphs. Then, it compares the feature vector to the training patients' feature vectors, sorting the training patients by similarity. The most similar patient's BA is returned to the clinician.

The nearest neighbour approach \cite{Cover1967-la} bears multiple advantages. First, the prediction is faster because no spot weight optimisation and dose calculations are needed. Second, each predicted BA was applied at least once in the clinic, rendering it as robust as the training patient's BAs. Third, there is no need to define the optimality of a beam arrangement because it is encoded in the dataset. Fourth, the model is capable of learning from datasets with varying number of beams, which is important for small clinics with heterogeneous datasets. Finally, the prediction is interpretable since the most similar patient(s) can be displayed to the clinician. Additional information from those patients such as optimisation objectives could also be re-used. We have not done so, but used the open source automatic spot weight optimisation algorithm JulianA for the spot weight optimisation \cite{Bellotti2023}.

The input features were chosen as follows based on discussions with professional treatment planners: The centre of mass (COM)  of the planning target volume (PTV) in the CT coordinates, PTV surface area, PTV volume and PTV sphericity. Further, the following features are calculated for each OAR: A boolean variable indicating whether the OAR has been contoured for this patient, the distance vector between the OAR and PTV COM in spherical coordinates and the minimum, mean and maximum distance between any point in the PTV and any point in the OAR. The OARs of interest are the brainstem, chiasm, cochleae (left and right), eyes (left and right), lacrimal glands (left and right), optic nerves (left and right), pituitary gland and hippocampi (left and right). In total, there are $97$ features. If an OAR is not contoured, the missing features are imputed with the median value of the other training patient's features. Patient similarity is calculated using the squared distance of the respective patients' feature vectors. Before calculating any features, a rigid registration is performed to homogenise length scales within the dataset.

The model is trained and tested on a dataset of $100$ patients, split into $81$ training and $19$ test stratified by the number of beams. The test set is the same dataset that was used in a prior study about automatic spot weight optimisation \cite{Bellotti2023}. All patients were treated with protons at our institute between 2013 - 2021. The inclusion criteria were to be treated at gantry two, presenting with a brain or eye tumour and being treated using three or four fields. We chose gantry two to facilitate clinical transition because its patients are treated exclusively with our in-house planning system FIonA, which offers best integration of the automatic spot weight optimiser JulianA. Head-and-neck patients were excluded because they are treated using standard beam arrangements according to our clinical protocol. This work aims to benefit especially institutes that do not have specialised beam arrangements for all indications, for example institutes to which wide ranges of indications are referred.

The dataset includes multiple indications on purpose to strengthen the utility of the model and provide additional value compared to indication-specific template solutions.
A rigid registration using a training patient as the reference is performed before calculating features to align the patient's CTs, which is necessary because our dataset includes both adult and pediatric patients.

Three beam arrangements (BAs) were selected using different methods: The reference beam arrangement was exported from our in-house treatment planning system (TPS) PSIPlan. This is the beam arrangement chosen manually by our dosimetrists for the actual treatment. The GAMBAS BA is generated by our model. The random neighbour BA (or simply random BA) is generated similarly to GAMBAS, but a random training patient was chosen instead of the most similar one. The random BA is needed to ensure that GAMBAS' performance is due to meaningful features rather than simply using treatment-approved angles.
A treatment plan is generated for each of the three methods using the open source spot weight optimiser JulianA \cite{Bellotti2023, code}. The resulting dose distributions are pseudonymised (labels A, B, C; different assignment for each patient) and handed to the chief treatment planner at our institute together with the CTs and the structure sets of each patient.

The treatment planner was tasked to assess the quality of each plan on a scale of 1-5 (1 - clearly unacceptable, 2 - rather unacceptable, 3 - acceptability unclear, 4 - rather acceptable, 5 - clearly acceptable). Further, the planner was asked to rank the plans and guess which method was used to generate each plan. All dose distribution analyses were executed in our in-house TPS FIonA.

\clearpage
\section{Results}
\begin{figure}
    \centering
    \includegraphics[width=0.75\textwidth]{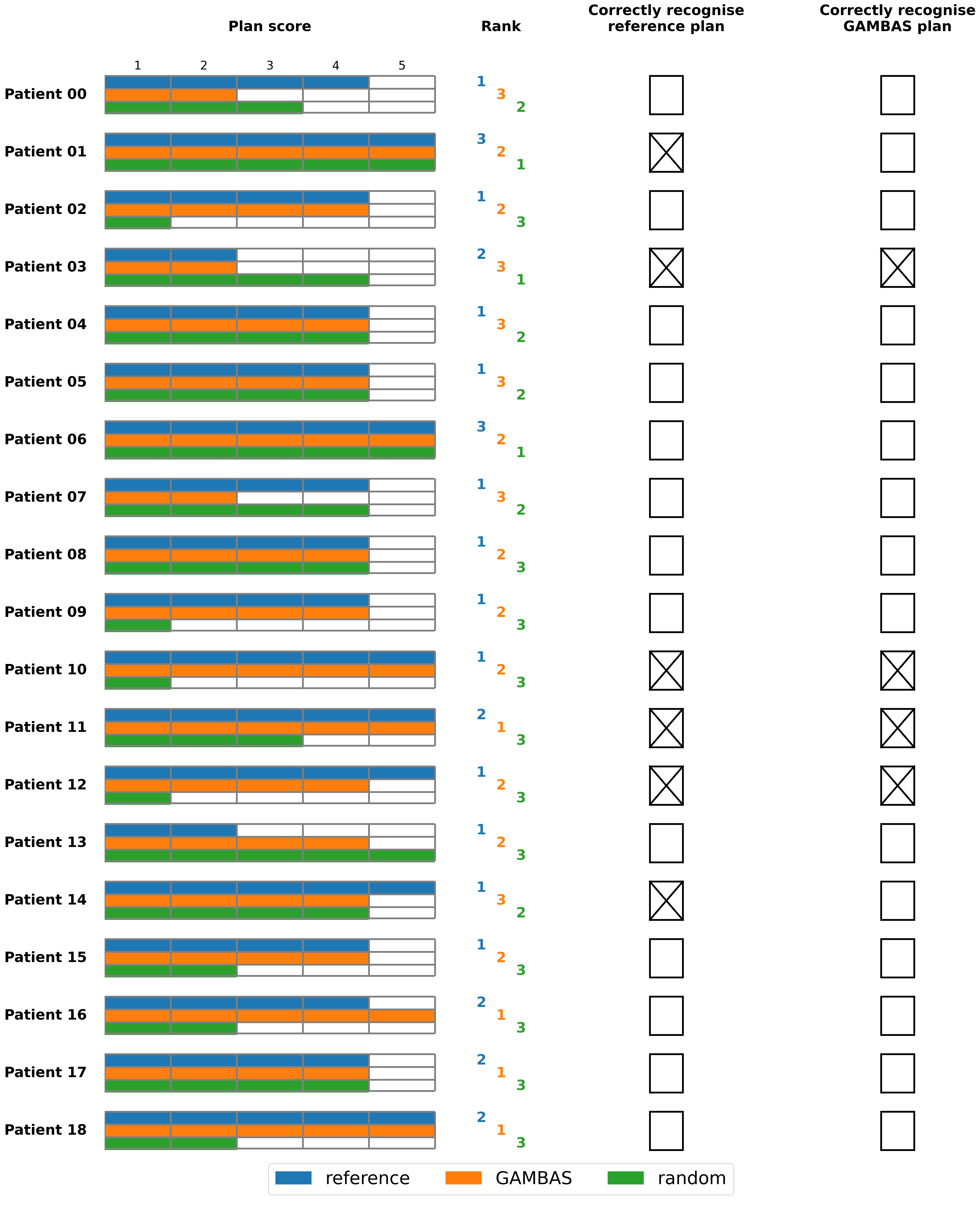}
    \caption{Summary of the plan assessment by an expert treatment planner. The first column indicates the quality of the plan (1 clearly unacceptable, 2 rather unacceptable, 3 unclear acceptability, 4 rather acceptable, 5 clearly acceptable). The second column section from the left indicates the ranking of the plans (1 indicates the most preferable plan, 3 the least preferable plan) and the third section denotes whether the expert identifies the reference and GAMBAS plans correctly.}
    \label{fig:summary}
\end{figure}
The results of the expert planner's assessment is depicted in Fig.~\ref{fig:summary}. The expert planner deemed $17~(89\%)$ of the reference plans, $16~(84\%)$ of the GAMBAS plans and $10~(53\%)$ of the plans generated with the random method acceptable (score $4$ or $5$). These numbers show that the reference and GAMBAS methods exhibit small differences in terms of acceptability rate, but each outperform the random method by more than $30\%$ points.

The most preferred plan was generated by the reference method for $12~(63\%)$, by GAMBAS for $4~(21\%)$ and by the random method for $3~(16\%)$ patients. Rank two was obtained by the reference method for $5~(26\%)$, by GAMBAS for $9~(47\%)$ and by the random method for $5~(26\%)$ patients. Finally, the reference method was deemed the least preferable for $2 (11\%)$ patients, the GAMBAS method for $6~(32\%)$ patients and the random method for $11~(58\%)$ patients. Concluding, the reference plans were preferred for the majority of patients and GAMBAS plans ranked second place for the majority of patients. The random plans are deemed the least preferential.

\begin{table}
    \centering
    \begin{tabular}{lrrr}
        \toprule
        \diagbox[width=6cm]{\textbf{Guessed method}}{\textbf{True method}} & reference & GAMBAS & random \\
        \midrule
        reference & 6~(32\%) & 12~(63\%) & 1~(~~5\%) \\
        GAMBAS & 10~(53\%) & 4~(21\%) & 5~(26\%) \\
        random & 3~(16\%) & 3~(16\%) & 13~(68\%) \\
        \bottomrule
    \end{tabular}
    \caption{Confusion matrix for the method recognition by the expert planner. Each row represents the guessed methods and each column the true method. Reading example: $10$ of the actual reference plans were mistaken as GAMBAS plans.}
    \label{tab:confusion}
\end{table}
The confusion matrix for the method recognition is shown in Tab~\ref{tab:confusion}. The diagonals signify the number of correct recognitions for each method. The randomly generated plan was identified correctly most often with $13~(68\%)$ correct identifications. The reference method was only correctly identified for $6~(32\%)$ and GAMBAS only for $4~(21\%)$ cases. The most frequent errors were to mistake the GAMBAS plan for the reference plan $12~(63\%)$ and vice versa $10~(53\%)$. Each of these confusions is more likely to occur than a correct guess for either the reference of GAMBAS method, suggesting similarity between the two methods.

\begin{figure}
    \begin{subfigure}[t]{0.45\textwidth}
        \centering
        \includegraphics[width=\textwidth]{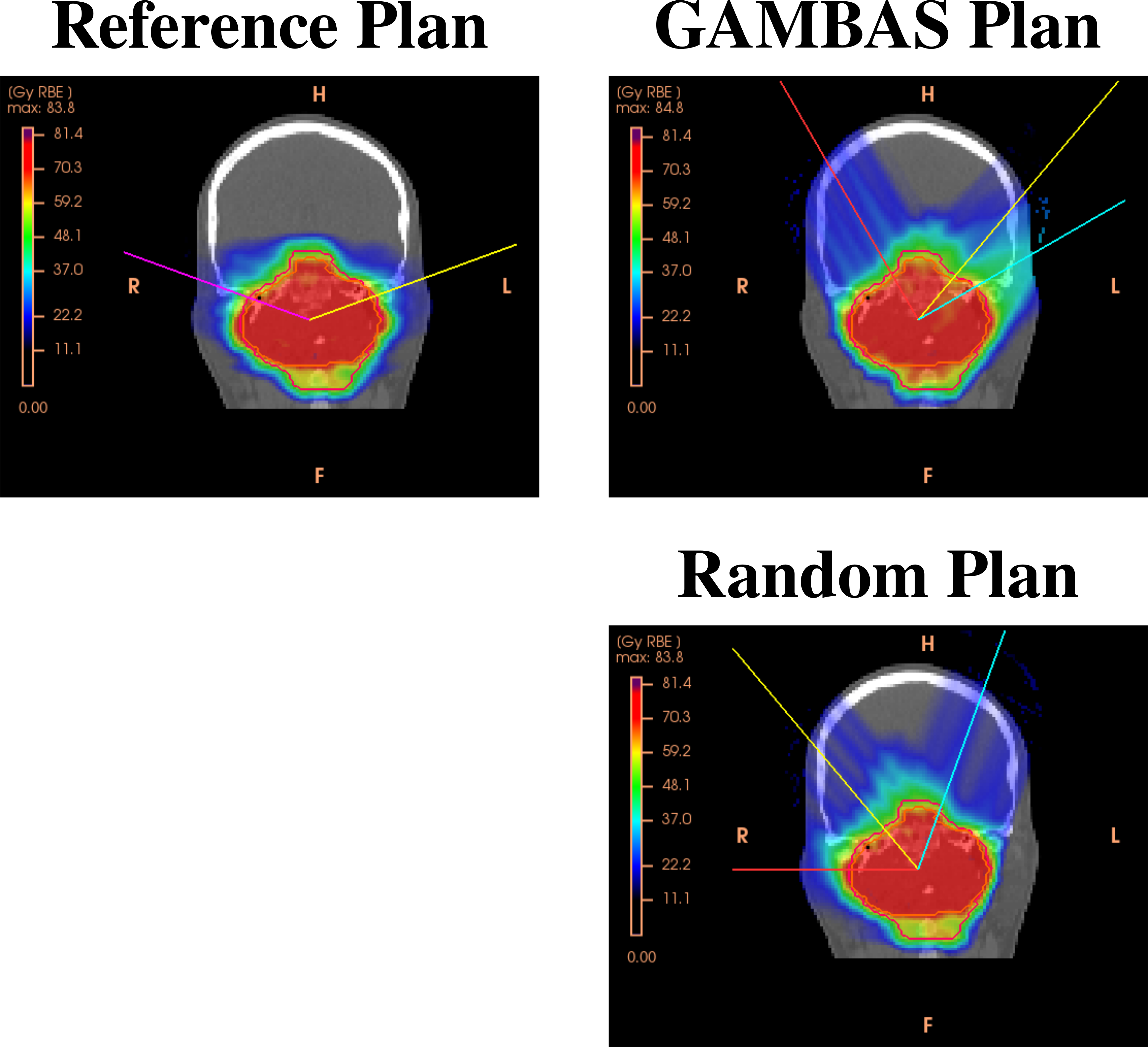}
        \caption{Coronal view on patient 07. It is one of the three cases where the GAMBAS plan was considered unacceptable.}
        \label{fig:test_07}
    \end{subfigure}
    \hfill
    \begin{subfigure}[t]{0.45\textwidth}
        \centering
        \includegraphics[width=\textwidth]{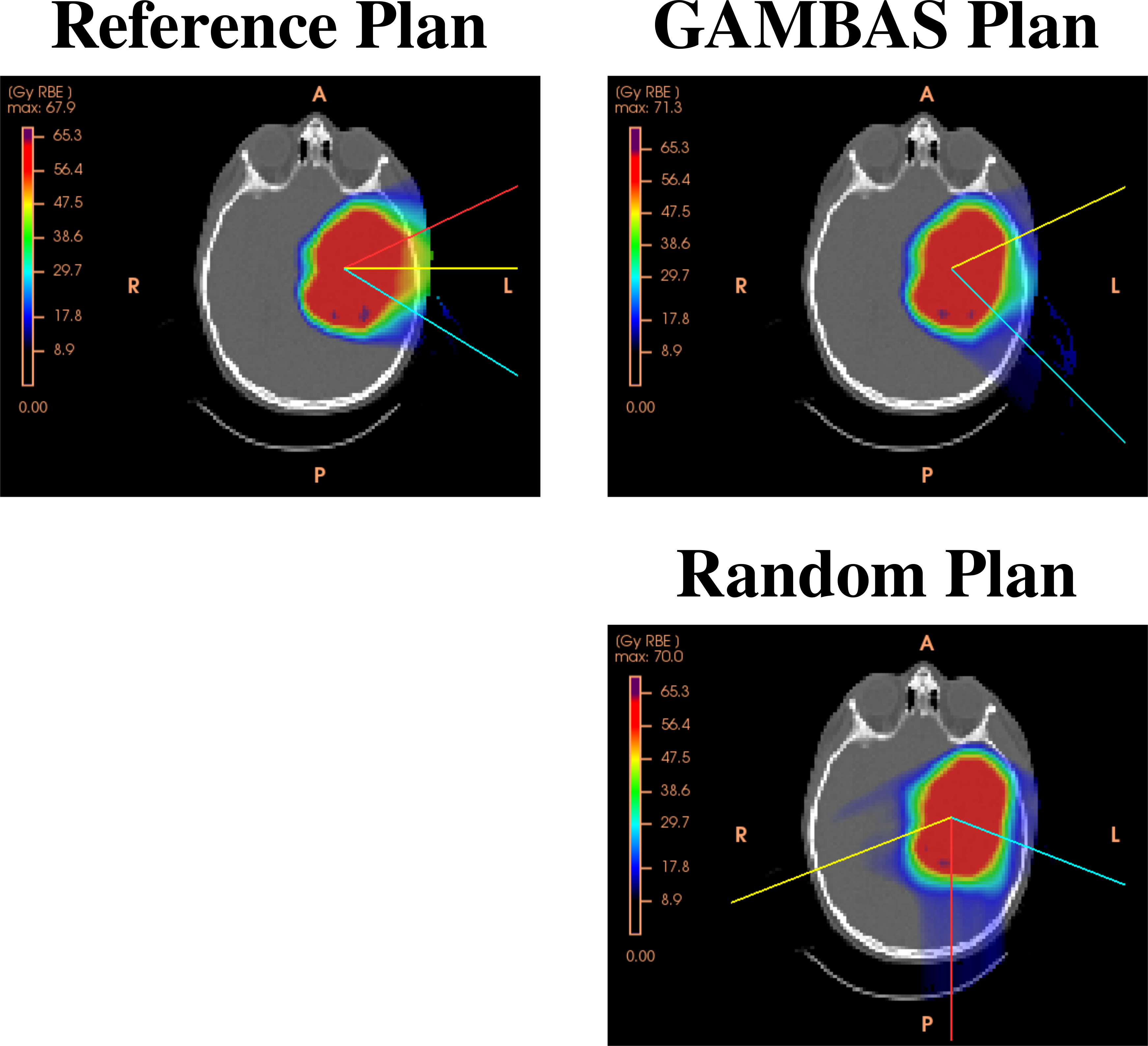}
        \caption{Axial view on patient 17. The GAMBAS plan was considered the best for this case.}
        \label{fig:test_17}
    \end{subfigure}
    \caption{Case studies. The coloured lines indicate the three beam directions. Reference plans use human-selected beam arrangements that were used for treatment. GAMBAS plans use model-predicted beam arrangements and random plans use a random training patient's beam arrangement. All spot weights were optimised using the open source JulianA algorithm. The in-house dose calculations were performed using our in-house planning system FIonA.}
\end{figure}
The results are concluded with a discussion of three example patients. Patient 07 (see Fig.~\ref{fig:test_07}) is one of the three cases where GAMBAS failed. For this patient, the GAMBAS and the random plan overdose the brain region unnecessarily by selecting beam angles that cross almost the entire cranium. The reference plan chose more flat gantry angles, keeping the dose to the brain minimal.\\
The GAMBAS plan is superior to both others for patient 17 (Fig.~\ref{fig:test_17}). The random beam arrangement does not cater to the unilateral position of the tumour, putting unnecessary dose to the right of the tumour, which is undesirable. Both the reference and the GAMBAS plans account for the unilaterality, but the reference plan chooses an almost co-planar arrangement, rendering the plan less robust than the GAMBAS plan that rotates one of the fields perpendicular to the other two. Dosimetrically, the reference and the GAMBAS plan are very similar.

Patient 03 is an interesting case because both the reference and GAMBAS plans were considered unacceptable, but the random plan is acceptable. This is because the PTV contours extend into air, which misguids the spot weight optimiser to cause severe hotspots at the patient's surface. This is a limitation of the spot weight optimiser and not due to the beam arrangement. The latter was considered acceptable for all plans.

The mean and standard deviation of the  time for beam arrangement prediction using GAMBAS are $1.1 \pm 0.2$~min (range: $0.8 - 1.6$min) on a single core of an Intel Xeon Gold 6152 processor. The main computational cost lies with the spot weight optimisation, which takes $4.1 \pm 0.7$~min (range $3.1 - 5.7$min) on an NVIDIA GeForce GTX 1080 graphics processing unit. The total planning time for the fully automatic treatment planning pipeline is $5.2 \pm 0.8$min (range $4.0 - 7.3$min).

\section{Discussion}
The results show that the GAMBAS-JulianA system is capable of generating clinically acceptable plans for $16~(84\%)$ patients, comparable to the $17~(89\%)$ acceptable plans obtained using human-selected beam angles with JulianA. The random treatment plans performed poorly, with (only $10~(53\%)$ acceptable plans), indicating their inferiority to both reference and GAMBAS plans.

Manual beam angle selection and automatic selection with GAMBAS are indistinguishable, with only $32\%$ of the reference plans and $21\%$ of the GAMBAS plans being identified correctly, whereas the random plans were correctly identified in $68\%$ of cases. This indicates that there are recognisable differences between the random neighbour approach, but not between the reference and GAMBAS methods.

These arguments provide evidence that GAMBAS beam arrangements are of a comparable quality to the reference plans and superior to random plans. Therefore, the selected features and similarity measures are meaningful and provide added value beyond reusing historically accepted treatment plans.

The GAMBAS-JulianA autoplanning system aims to achieve plans of a comparable quality to human-created plans within the tight time frame and limited amounts of data typical for clinical settings. Requiring less than \SI{8}{min} to create a full deliverable treatment plan, the GAMBAS-JulianA system surpasses all existing autoplanning tools in terms of speed, to the best of our knowledge.\\
Further, this system is suitable for low-resource environments such as research facilities or small clinics. JulianA does not require any training data at all and GAMBAS can be trained in a short time and on a conventional workstation without powerful hardware, using a heterogeneous dataset with a variety of indications, tumour locations, shapes and sizes as well as a varying number of beam angles.

The main advantage of the system, aside from speed, is its simplicity. Once trained, no manual annotations or special preprocessing steps are required compared to manual treatment planning. Only the CT, structure contours, prescribed dose and OAR dose threshold values are needed.

Further research should explore generalising GAMBAS to other treatment sites such as lung, abdomen or pelvis, and data of other institutes. Collaborations in this area are encouraged.\\
Though manual beam arrangements are still preferred to GAMBAS beam arrangements for the majority of patients in this dataset, further research could explore whether the beam arrangement predicted by GAMBAS would accelerate existing beam arrangement optimisation algorithms by providing an initial guess, or how much manual adjustment would be needed to achieve human-level performance.\\
Currently, the similarity-metric is crafted based on expert reasoning. However, future research should investigate even better similarity metrics.

Further research should complement this study before GAMBAS can be used for treating patients. The model's capabilities need to be investigated on a larger test set to improve the statistical meaningfulness of the validation results. Further, the next validation study should rely on more than a single expert for the plan review in order to address and quantify the inter-observer variability that is unavoidable in treatment planning \cite{Berry2016-ew, Nelms2012-ri}. Finally, robustness analysis was beyond the scope of this pilot study. Further research should employ robust optimisation and conduct a thorough robustness analysis to investigate the hypothesis that GAMBAS beam arrangements are equally robust as dosimetrist-chosen arrangements due to the fact that each GAMBAS arrangement has been used for treatment before.

The GAMBAS model has already been used to accelerate research endeavours at our institute, especially to accelerate validation studies of machine learning models. GAMBAS produced beam angles for an MR-to-CT model \cite{li2024}, significantly reducing the time required for assessing the model's dosimetric impact. We intend to use GAMBAS more often for ongoing and future projects to improve the efficiency and reproducibility of research at our institute even more.

The GAMBAS-JulianA system also promises interesting applications for clinical environments, mainly as a quality assurance and patient selection tool. Given the low resource requirements of the system, an autoplan could be generated for every patient suffering from compatible tumours. The autoplan would then serve as a baseline treatment plan and automatic QA tools would raise warnings if the target coverage or achieved OAR sparing deviates too much from the values achieved by the autoplan, contributing to an even more reliable plan approval workflow.\\
Automatic treatment planning has been investigated as a resource-efficient approach for patient selection for proton therapy \cite{langendijk2013, hytonen2022}. Previously published results focus on knowledge-based planning methods to automate the spot weight optimisation, but use the same fixed beam arrangement for each patient \cite{hytonen2022}. GAMBAS could contribute to such patient selection pipelines by providing a patient-specific beam arrangement, rendering the selection even more individualised.

\section{Conclusions}
GAMBAS, a novel fast knowledge-based beam angle prediction model, was developed and analysed. GAMBAS beam arrangements are generated in less than \SI{2}{min} and cannot reliably be distinguished from human-chosen beam arrangements and achieve a similar acceptance rate than the latter. GAMBAS is already accelerating research at our institute and we plan to deploy it as part of an assistant tool for QA and patient selection.

\section{Highlights}
\begin{itemize}
    \item A novel method for automatic beam angle selection in proton therapy is presented.
    \item The method obtains a solution in $\leq \SI{2}{min}$.
    \item It is based on interpretable anatomical features and patient similarity.
    \item The auto plans are deemed acceptable for treatment for $16 (84\%)$ cases.
    \item The model is ready for quality assurance, patient selection and research use cases.
\end{itemize}

\section{Code Availability}
The code for this study is published as open source software \cite{gambas_code}.

\section{Declaration of Generative AI and AI-assisted technologies in the writing process}
During the preparation of this work the author(s) used ChatGPT in order to reformulate and shorten the abstract, Introduction and Discussion sections. After using this tool/service, the author(s) reviewed and edited the content as needed and take(s) full responsibility for the content of the publication.

\printbibliography

\end{document}